\documentclass[reprint,twocolumn,twoside,amsmath,amssymb,showpacs,aps,nofootinbib,prd]{revtex4-1}

\usepackage{graphicx}
\usepackage{epsfig}
\usepackage{amssymb}
\usepackage{dcolumn}
\usepackage{bm}
\usepackage[unicode]{hyperref}
\usepackage{bookmark}

\usepackage{verbatim}
\usepackage[utf8]{inputenc}

\newcommand{\ba}{\begin{array}}
\newcommand{\ea}{\end{array}}
\def\br{\begin{eqnarray}}
\def\er{\end{eqnarray}}
\def\be{\begin{equation}}
\def\ee{\end{equation}}

\def\({\left(}
\def\){\right)}

\begin{document}

\title{$R_{e^+ e^-}$ and an effective QCD charge}
 
\author{J.~D.~Gomez$^1$ and A.~A.~Natale$^{1,2}$}
\affiliation{$^1$Centro de Ci\^encias Naturais e Humanas, Universidade Federal do ABC, 09210-170, Santo Andr\'e - SP, Brazil \\
$^2$Instituto de F\'{\i}sica Te\'orica, UNESP, Rua Dr. Bento T. Ferraz, 271, Bloco II, 01140-070, S\~ao Paulo - SP, Brazil}              

\date{\today}

\begin{abstract}

We consider the electron-positron annihilation process into hadrons $R_{e^+e^-}$ up to $\mathcal{O}(\alpha_{s}^{3})$ and we adopt the smearing method suggest by Poggio, Quinn and Weinberg to
confront the experimental data with theory. As a theoretical model we use a QCD coupling constant frozen in the low energy regime, where this coupling can be parameterised in terms of an effective dynamical gluon mass ($m_g$) which is determined through Schwinger-Dyson equations. In order to find the best fit between experimental data and theory we perform a $\chi^2$ study,
that, within the uncertainties of the approach, has a minimum value when $m_g/\Lambda_{QCD}$ is in the range $1.2 \, - \, 1.4$. These values are in agreement with other phenomenological determinations of this ratio and lead to an infrared effective charge $\alpha_s(0) \approx 0.7$. We comment how this effective charge may affect the global duality mass scale that indicates the frontier between perturbative and non-perturbative physics.

\end{abstract}


\maketitle

\section{Introduction}
The standard model is one of the most successful theories of the last century, whose physical quantities are computed at the loop level with high precision showing remarkable agreement with the high energy experimental data.
On the other hand,  the theory that explains the strong interaction, known as Quantum Chromodynamics (QCD), is governed by an asymptotically free gauge field theory involving elementary
quark and gluon fields at high energy or short distances. This behaviour at high energies is amenable to perturbative theory calculations using  Feynman diagrams.  This can be used to calculate physical mass-shell process.
At low energy there is no justification for a perturbative treatment of QCD.  In this regime non-perturbative approaches come into play. 

A non-perturbative approach for QCD is provided by the Schwinger-Dyson equations (SDE), whose study has revealed relevant progress in the recent years, most of them related to the disappearance
of infrared (IR) divergences when the theory is resumed in a particular gauge invariant scheme, known as pinch technique \cite{pinch}.
The infrared QCD coupling turns out to be IR finite when gluons develop a dynamically generated mass ($m_g$) in this non-perturbative approach. This point was first demonstrated in Ref.\cite{cornwall} and was also discussed at length in Refs.\cite{papa1,papa2,papa3,papa4,papa5,papa6,papa7,papa8} among many other references of this group. These results were checked
independently by different lattice simulations \cite{lat1,lat2,lat3,lat4,lat5,lat6,lat7,lat8}, has been studied in different approaches \cite{sor1,sor2,sor3,sor4}, and is, step by step, being accepted as a cure of the infrared QCD divergences. 
The finitude of the QCD coupling constant can be related to a non-perturbative IR fixed point, which is a property of dynamical mass generation in non-Abelian theories \cite{us1}. The phenomenological consequences of such IR finite coupling, or non-perturbative fixed point,
have been discussed in Ref.\cite{us2}, and recently we have discussed how this non-perturbative fixed point can change the local minimum of a renormalisation group improved effective potential \cite{us3}. This change of
minimum state may produce noticeable modifications in the physical properties of the model studied in Ref.\cite{us4}.

From the phenomenological point of view the theoretical results leading to a finite QCD coupling is most than welcome, since many models describing strong interaction physics at low
energy or small transferred momenta make use of an IR finite moderately small coupling constant. We present in the sequence a partial list of model calculations using an IR finite coupling constant: 1) The description of jet shapes observables requires an IR coupling equal to $0.63$ \cite{Webber}, 2) Quarkonium potential models use an IR coupling of order 
$0.6$ \cite{quarko}, 3) The theoretical ratio $R_{e^+e^-}$ can fit the experimental data with an IR coupling approximately $0.8$ \cite{Matt}, 4) Calculation of quarkonium fine
structure in the framework of the background perturbation theory require a coupling as low as $0.4$ \cite{dezset}, 5) QCD-inspired models describing total hadronic cross sections
make use of an IR coupling of the order $0.5$ \cite{block}, 6) The experimental data on the unpolarised structure function of the proton is fitted with a coupling
constant in the range $0.4\, - \, 0.56$ \cite{court}. Besides the phenomenological indications, there are also several theoretical hints that the QCD coupling does not increase 
abruptly at low energies, and this fact may explain results claiming that the frontier between perturbative and non-perturbative physics may occur at relatively small 
momenta \cite{levin}. Of course, many of these calculations are based on models that cannot be fully connected to QCD, and even when related to QCD they may be, for example, renormalisation
scheme dependent, but they provide one hint about what we can expect for the IR value of the strong coupling constant. 

It should be remembered that
a frozen coupling constant of the order that we described in the previous paragraph may contradict what is known about chiral symmetry breaking in QCD \cite{ne35,ne36}.
It is indeed a problem in the realm of Schwinger-Dyson equations
to obtain the right chiral parameters for the value of the quark condensate, $f_\pi$ and other quantities when the gluon acquires a
dynamically generated mass and the coupling freezes in the infrared. Possible solutions in this approach, invoking non-perturbative QCD aspects, were pointed out
only in the last years \cite{ne37,ne38}. The calculation of Ref.\cite{ne37} is consistent with lattice data 
showing that the chiral and confinement transitions happens at the same temperature \cite{ne39}, and within this model we can obtain reasonable values for the chiral parameters \cite{ne40}.
Fortunately the problems raised by Peris and de Rafael \cite{ne35} do not lead to a ``no-go" theorem, and the solution of Ref.\cite{ne37} is one possible way to evade these difficulties.

In this work we will study the ratio $R_{e^+e^-}$:
\be
\label{eq:1}
R_{e^+e^-}(s)\equiv\frac{\sigma(e^+e^-\rightarrow hadrons)}{\sigma(e^+e^-\rightarrow \mu^+ \mu^-)},
\ee
as a tool to determine the infrared value of an effective QCD charge, which is related to the ratio $m_g/\Lambda_{QCD}$, where $\Lambda_{QCD}$ is the QCD characteristic scale. A similar kind of analysis has already been performed
in the known work of Mattingly and Stevenson \cite{Matt} in the context of Optmised Perturbation Theory (OPT) and in Ref.\cite{milton} in the context of Analytic Perturbation
Theory (APT). Here we will differ from them using an  effective QCD charge obtained from Schwinger-Dyson equations, verifying that the experimental
data is fitted only by  a narrow range of $m_g/\Lambda_{QCD}$ values. 
We recall that $R_{e^+e^-}(s)$ can be calculated perturbatively, and can be developed as a power series in the QCD coupling, and in terms of the parameter
\be
a(s)\equiv\alpha_s(s)/\pi,
\label{eq1a}
\ee
which, in our case, will be improved using the effective charge obtained through the SDE in an gauge invariant way using the Pinch technique \cite{pinch}.

The organisation of this work is the following: In Section II we discuss the hadronic cross section $R_{e^+e^-}(s)$ in both partonic level and with massive quarks, and after that we use the non-perturbative effects in the QCD coupling constant to calculate the behaviour of ratio  $R_{e^+e^-}(s)$. In Section III the theoretical calculation is compared with the experimental data using smeared functions as
proposed by Poggio, Quinn and Weinberg \cite{pqw}, obtaining results that confirm the IR freezing of the QCD coupling constant. In Section IV we use the concept of global duality and discuss how this effective charge affects the scale that indicates the matching between the
perturbative and non-perturbative physics.
In Section V we draw our conclusions.

\section{Hadronic cross section \texorpdfstring{$R_{e^{+}e^{-}}(s)$}{Lg}}

The $e^+e^-$ annihilation into hadrons is one of the most important processes for testing the theory of strong interaction.  This process provides a fundamental QCD test, supplying evidence for the existence of colour \cite{EllisWebber}.
At lowest order the total hadronic cross section is obtained by simply summing over all kinematically accessible flavours and colours of quark-antiquark pairs,
\be
\label{eq:ReeParton}
R_{e^+e^-}(s)\equiv\frac{\sigma(e^+e^-\rightarrow hadrons)}{\sigma(e^+e^-\rightarrow \mu^+ \mu^-)} = 3\sum_{i}^{n_f}q_{i}^{2}.
\ee
Here the $q_{i}$ denote the electric charges of the different flavours of quarks, and this is recognised as the parton model result.

Real and virtual gluon corrections to this basic process (\ref{eq:ReeParton}) will generate higher-order contributions to the perturbative series.  The second and higher-order corrections in perturbation theory were computed
a long time ago in the zero quark mass limit in Ref.\cite{correctionRee}, and can be expressed order by order in a perturbative series 

\be
\label{eq:ReePartonSerie}
R_{e^+e^-}(s)=3\sum_{i}^{n_f}q_{i}^{2}(1+\mathcal{R}(s)),
\ee
where $\mathcal{R}$ has the form
\be
\label{eq:Rform}
\mathcal{R}(s)=a(1+r_1 a + r_2 a^2+\cdots),
\ee
and depends upon a single kinematic variable $s\equiv Q^2$, the c.m. energy.  As shown in Eq.(\ref{eq1a}) $a$ is defined by the QCD coupling constant over $\pi$. The coefficients of the Eq. \eqref{eq:Rform} depend of the renormalisation scheme. 
In the modified Minimal Subtraction scheme ($\overline{MS}$)
the coefficients were computed in Ref.\cite{correctionRee}.
\begin{gather}
\label{eq:r1}
r_1 = 1.986 - 0.1153n_f^{2}  \\
r_2 = -6.637-1.200 n_f - 0.00518 n_f^2, 
\label{eq:r2}
\end{gather}
 where $n_f$ is the number of flavours.\\ 

We will follow closely the pioneering work about OPT by Mattingly and Stevenson \cite{Matt}, but instead, we will consider  the non-perturbative approach to the QCD coupling
constant obtained through SDE.  The approach described above is valid for massless quarks, $m_q=0$. To include quark mass effects we use the approximate result \cite{Politzer,Matt,pqw}
\be
\label{eq:ReeQmass}
 R_{e^+ e^-}(Q^2)=3\sum_{i} q_{i}^2 \frac{v_i}{2}(3-v_{i}^2)[1+g(v_i)\mathcal{R}],
\ee
with the sum over the active quark flavours that are above threshold, i. e. those with masses lower that $Q/2$ ($m_i < Q/2$), and 
\begin{equation}
 \label{eq:gv}
 \begin{gathered}
  v_i=\Big(1-4\frac{m_{i}^2}{Q^2}\Big)^\frac{1}{2},\\
  g(v)=\frac{4\pi}{3}\Big[\frac{\pi}{2v}-\frac{3+v}{4}\Big(\frac{\pi}{2}-\frac{3}{4\pi}\Big)\Big].
 \end{gathered}
\end{equation}
Here $v_i$ represents the quark velocity, so that $v_i=0$ corresponds to the heavy quark threshold, and the results for massless quarks are recovered in the relativistic limit $v_i\rightarrow 1$.

For the theoretical calculation of $R_{e^+e^-}(s)$ we shall take $\mathcal{R}(s)$ giving by Eq. \eqref{eq:ReePartonSerie} and \eqref{eq:Rform} with the coefficients $r_1$ and $r_2$ 
given by Eqs.\eqref{eq:r1} and \eqref{eq:r2} respectively, but assuming an expansion in terms of the non-perturbative QCD coupling constant obtained from QCD Schwinger-Dyson equations
in the pinch technique approach. A quite general expression for this coupling is given by \cite{papa2}
\be
\alpha_s (k^2)=\Bigg[4\pi {\beta_0}\ln\Bigg(\frac{k^2+f\big(k^2,m^2_g(k^2)\big)}{\Lambda_{QCD}^{2}}\Bigg)\Bigg]^{-1},
\label{eq:2}
\ee
where the function $f\big(k^2,m^2(k^2)\big)$ is determined as a fit to each specific value of the dynamically generated effective gluon mass $m_g$.
For simplicity we have adopted the following expression for this coupling 
\be
\alpha_s (Q^2)=\Bigg[4\pi {\beta_0}\ln\Bigg(\frac{Q^2+\rho m_g^{2}(Q^2)}{\Lambda_{QCD}^{2}}\Bigg)\Bigg]^{-1},
\label{eq:alpha}
\ee
where $\beta_0=(11N-2n_f)/48\pi^2$, $\Lambda_{QCD}\equiv \Lambda$ is the characteristic QCD scale, and the function $m_g^{2}(Q^2)$ represents the dynamical gluon mass given by \cite{agui}
\be
m_g^2(Q^2) \approx \frac{m_{g}^{4}}{Q^2+m_{g}^{2}}.
\label{eq:mg_k}
\ee            
The advantage of Eq.(\ref{eq:alpha}) is that we can vary the IR $m_g$ value without the need of solving the coupled SDE for the gluon propagator, and as long as
we consider the $3<\rho<4$ interval we accommodate the early Cornwall's result \cite{cornwall}, the ones of Ref.\cite{papa2} without noticeable
differences, and this range is also consistent with the phenomenological values obtained in Ref.\cite{court}. It is important to note that at high energies we recover the usual perturbative strong coupling.

The effective QCD charge discussed in the previous paragraph has been obtained in one specific scheme (SDE and Pinch Technique) leading to a particular derivation of the non-perturbative QCD effective coupling.
Although this scheme is gauge invariant \cite{pinch} it may be claimed that we could have different definitions for the non-perturbative QCD coupling, and argue why we should consider Eq.(\ref{eq:alpha}) as
representative of the actual behaviour of the QCD charge. About this we can first say that Eq.(\ref{eq:alpha}) matches with the perturbative QCD coupling at high momenta. Secondly, 
lattice data and SDE solutions are clearly pointing to the existence of a dynamical mass scale for the gluon propagator, and if this is true we can prove the existence of a non-perturbative IR fixed point, i.e. an IR frozen coupling
constant \cite{us1}. Therefore, Eq.(\ref{eq:alpha}) certainly matches the expected ultra-violet (UV) and IR behaviours of the QCD coupling constant no matter the scheme used to determine this effective charge, and we shall assume that in our calculation
we can replace the perturbative coupling by this effective one.

Going back to the $R_{e^+e^-}(s)$ ratio, one interesting issue is to compare the effects of quark masses at low energy on this ratio when we consider the effective QCD coupling constant $\alpha_s$. In other words, we take both equations
Eq. \eqref{eq:ReePartonSerie} and Eq. \eqref{eq:ReeQmass} at order $\mathcal{O}(\alpha_s)$, and compare them.  So these equations becomes

 \begin{gather}
  \label{eq:NPReeMass}
   R_{e^+ e^-}(Q^2)=3\sum_{i} q_{i}^2 \frac{v_i}{2}(3-v_{i}^2)\Big[1+g(v_i)\frac{\alpha_s(Q^2)}{\pi}\Big],\\
   R_{e^+e^-}(Q^2)=3\sum_{i}^{n_f}q_{i}^{2}\Big(1+\frac{\alpha_s(Q^2)}{\pi}\Big),
   \label{eq:NPRee}
 \end{gather}
where $\alpha_s(k^2)$ is given by Eq. \eqref{eq:alpha}. We take standard values for the current-quark masses \cite{PDG}: $m_u=2.4$ MeV, $m_d=4.9$ MeV, $m_s=100$ MeV, $m_c=1.3$ GeV and  $\Lambda=300$ MeV.

\begin{figure}[htbp]
\setlength{\epsfxsize}{1.0\hsize} \centerline{\epsfbox{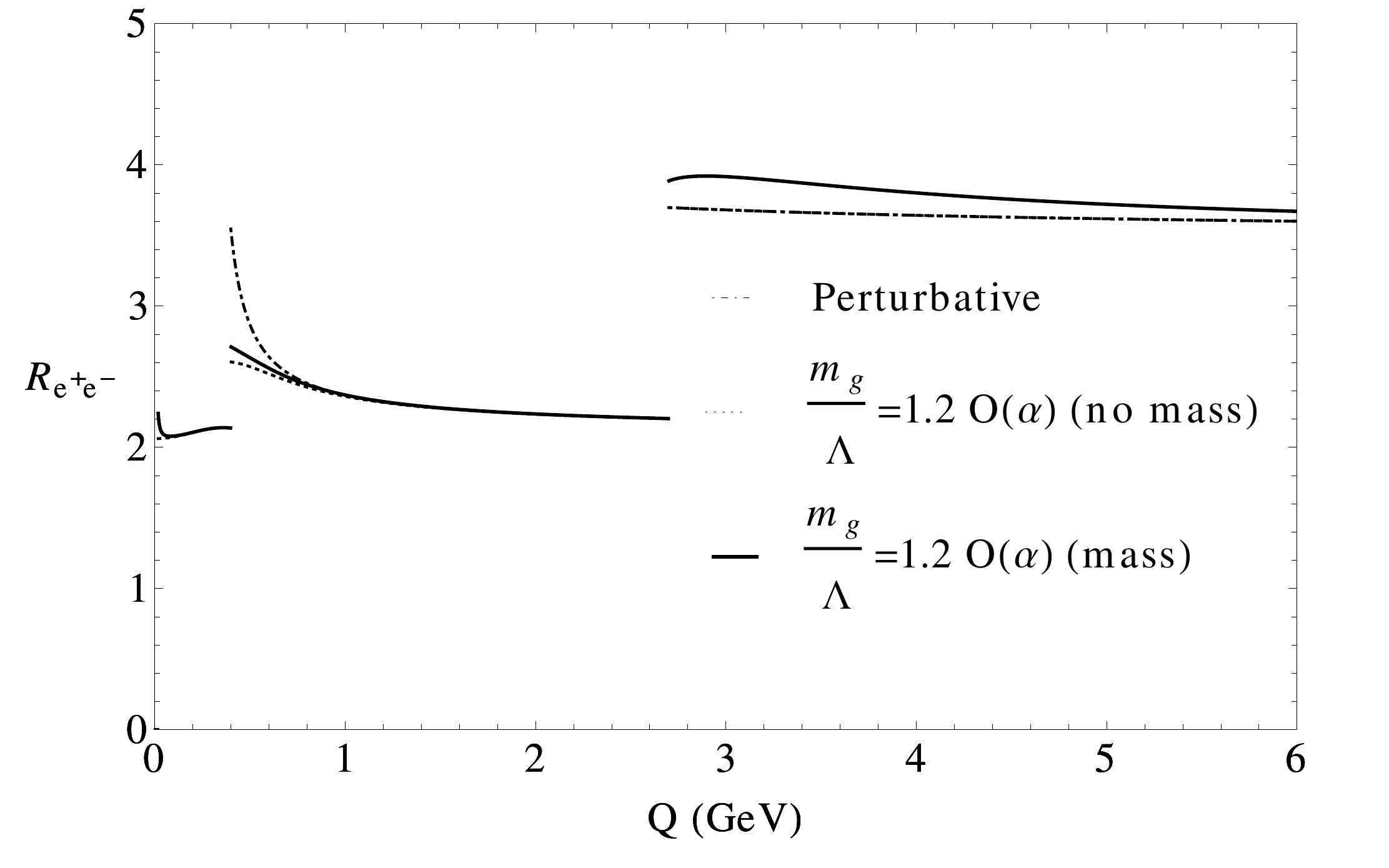}}
\caption[dummy0]{The one-loop theoretical ratio $R_{e^+e^-}$ behaviour with the non-perturbative coupling of Eq.\eqref{eq:alpha} with $\rho =4$ and $m_g/\Lambda=1.2$. The mass effects on the $R_{e^+e^-}$ is given by Eq.\eqref{eq:NPReeMass} (solid line),
                 while the Eq. \eqref{eq:NPRee} is represented by the dashed line. The  calculation with the perturbative coupling constant is also shown by the dot-dashed line.}
  \label{fig:TheoRee}
\end{figure}   

From Fig.\eqref{fig:TheoRee} we can see that the mass effect on the ratio $R_{e^+e^-}$ is negligible for light quarks, and become important with the opening of the charm threshold. Note that not only the freezing in the QCD coupling
constant (Eq.\eqref{eq:alpha}) was considered into Eqs.\eqref{eq:NPReeMass} and \eqref{eq:NPRee}, but the perturbative coupling constant has been considered into Eq.\eqref{eq:NPRee} as well, and this one is represented by the dotted-dashed line
in Fig.\eqref{fig:TheoRee}. It seems that the mass effect, when we use the IR finite charge, is not so strong in the low energy regime (below $1$ GeV)  in comparison
with the pure perturbative calculation.
  
We have considered 
the expression for $R(s)$ in the ${\overline{MS}}$ and the same happens for the quark masses. The mass effect is
only relevant for heavy quarks, does not affect the main region of R(s) that we study when the coupling is infrared finite, and the mass effect was not included
in the final result. However we do use a different scheme to describe the
infrared finite coupling!
It has been argued that the Green's functions obtained through the combination of the pinch technique with the
background field method (PT-BFM) are gauge invariant and renormalization group independent \cite{ne41} (and so the coupling constant \cite{ne42}),
i.e. they are independent of any renormalization mass $\mu$. This means that in principle we could
obtain a coupling that would be independent of ambiguities in its determination. However this is not the case. The Schwinger-Dyson
equation (SDE) for the gluon propagator, from where it is obtained part of the information leading to the infrared coupling, has to be solved
imposing that the non-perturbative propagator is equal to the perturbative one at some high-energy scale $(\mu )$, or comparing
the SDE propagator to the lattice data. After obtaining the QCD propagators we can determine the $\mu$ independent coupling through one specific relation
of two point correlators. This procedure is not unambiguous and contains all the numerical uncertainty related to this
specific calculation. We may argue that the mass generation mechanism in the PT-BFM approach minimizes the vacuum energy \cite{cornwall} and should reflect a scheme independent quantity, however the full procedure to obtain
the infrared finite coupling contains truncation and numerical approximations that are hard to be estimated at the present level of the SDE solutions.
As a matter of completeness we recall that another possible determination of the IR-behavior of $\alpha_s$ is based on Light-Front holographic 
QCD \cite{ne43}. The only assumption that underlies Light-Front holographic QCD is that QCD is conformal in the IR, in a 
procedure claimed to be renormalisation scheme independent \cite{ne44}. Similarly to the above discussion, the determination of the coupling uses a particular definition, which is called the $g_1$ scheme, that may introduce uncertainties in the same way as the PT-BFM approach. The $g_1$ scheme can be related to other couplings, as the one assumed in our work, using fundamental QCD relations, and these relations suggest that, once transformed to the one discussed
here, the coupling freezes near $0.7$ \cite{ne45}, i.e. the same value found by us.

\section{Smearing $R_{e^+e^-}(s)$}

The comparison of the theoretical prediction with the experimental data for $R_{e^+ e^-}$ is not possible rigorously, because there is no direct correspondence between the perturbative quark-antiquark thresholds and the hadronic thresholds and resonances
of the data. The only way to do this is using the smearing method proposed by Poggio, Quinn and Weinberg \cite{pqw}. Although this method was thought to be used at high
energies, in the Refs. \cite{Matt} and \cite{milton} it was used at low energies in different contexts and with interesting results. 

\begin{figure}[htbp]
\setlength{\epsfxsize}{1.0\hsize} \centerline{\epsfbox{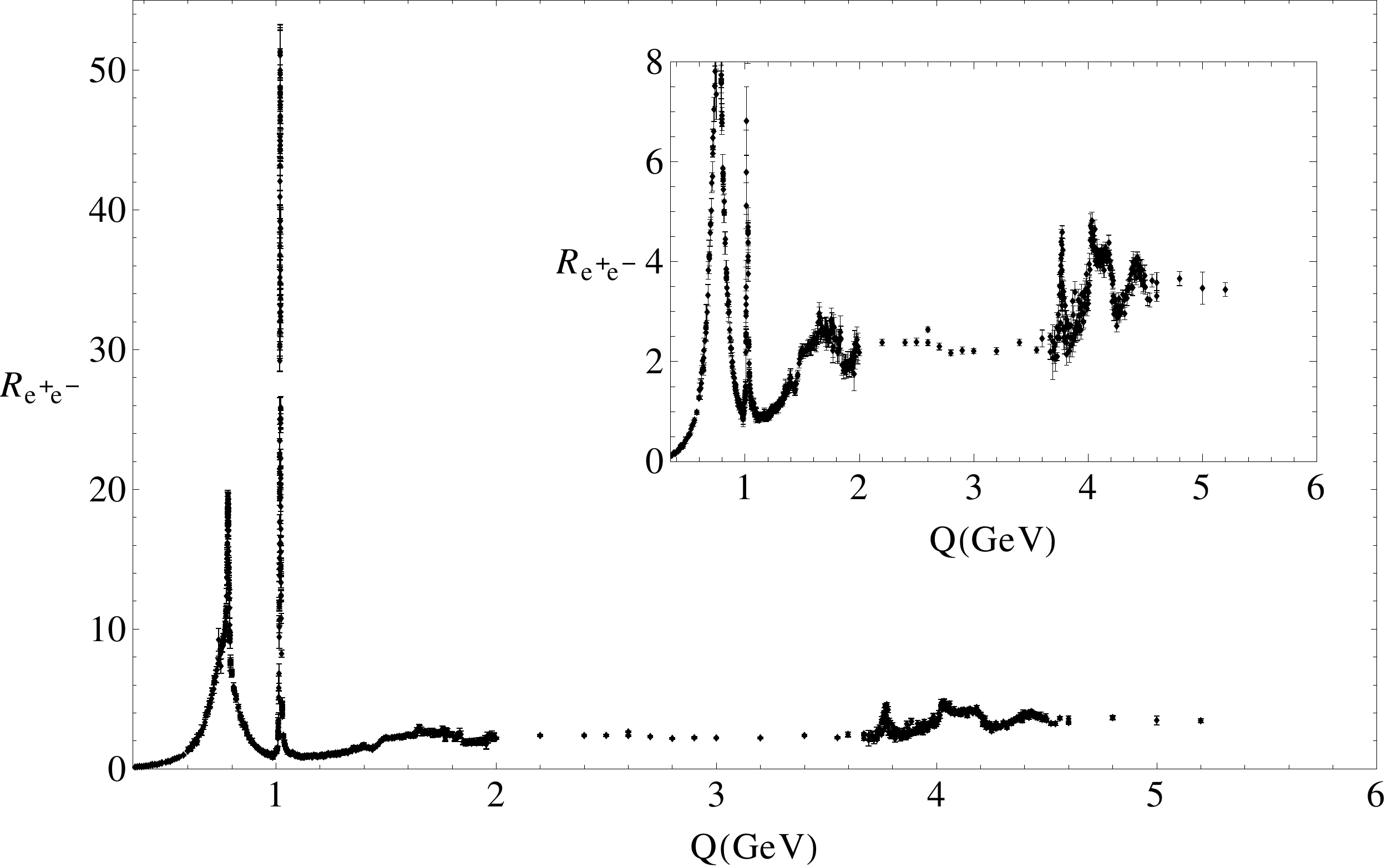}}
\caption[dummy0]{World data to ratio $R_{e^+ e^-}(s)=\sigma(e^+e^-\rightarrow \mathrm{hadrons},s)/\sigma(e^+e^-\rightarrow \mu^+\mu^-,s)$. $\sigma(e^+e^-\rightarrow \mathrm{hadrons},s)$ is the experimental cross section corrected
for initial state radiation and electron-positron vertex loops, $\sigma(e^+e^-\rightarrow \mu^+\mu^-,s)=4\pi\alpha^2(s)/3s$.  Data errors are total below 2 GeV and statistical above 2 GeV.}
  \label{fig:ExpRee}
\end{figure}

The experimental data that we are going to use has been taken from Particle Data Group Ref.\cite{PDG}.  Of particular interest to us is the region from $Q=0$ up to $6$ GeV.
The data and errors are shown in Fig.\eqref{fig:ExpRee}.  In that figure we have done a zoom to better show the resonances in the region of interest to us.

\begin{figure}[ht!]
\setlength{\epsfxsize}{1.0\hsize} \centerline{\epsfbox{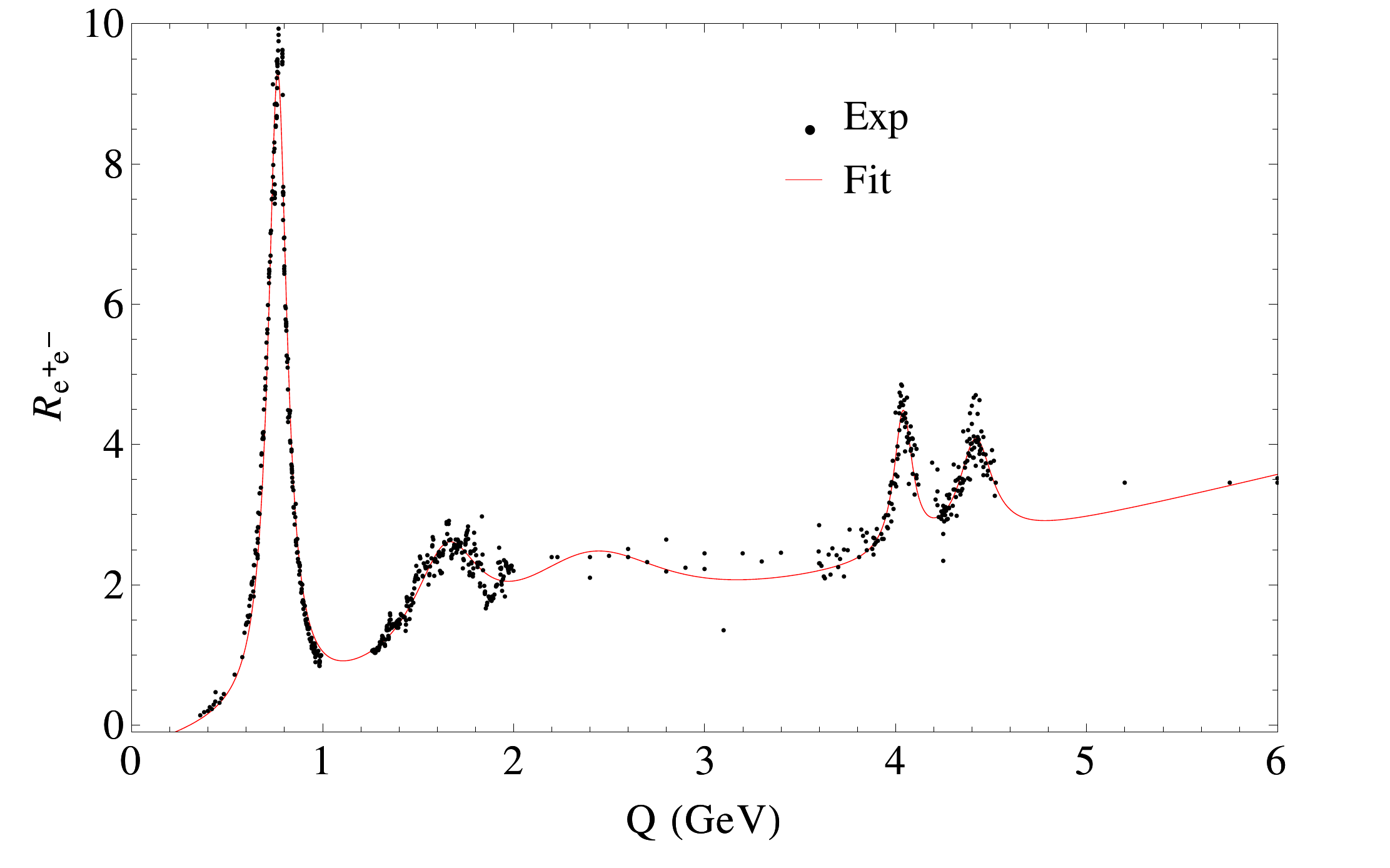}}
\caption[dummy0]{Compilation of experimental $R_{e^+ e^-}$ data (excluding narrow resonances). The red solid line represents the fit of Eq.\eqref{fitajuste}.}
  \label{fig:ExpReeFit}
\end{figure}

Figure \eqref{fig:ExpReeFit} shows our data compilation, up to 6 GeV. The red solid line represents our fit, where the quite narrow resonances $\rho,\,\phi,\,J/\psi,\,\psi(3686),\,\psi(3770)$,
as assumed in Ref.\cite{Matt}, were not included into the data compilation. The data go well beyond the $b$ quark threshold, but they have no real effect on the results that we shall
present. In the fit of the Fig.\eqref{fig:ExpReeFit} are included four resonances, the $\omega,\,\rho^{\prime},\,\psi(4040),\,\psi(4415)$. The red curve was obtained with the 
following fit:
\be
\sum_{i}^{4}\frac{A_i B_i}{(Q^2-M_i^{2})^2+A_i}+C+D Q^2,
\label{fitajuste}
\ee
where $M_1=0.781,\,M_2=1.65,\,M_3=4.04,\,M_4=4.42$ GeV$^{2}$ and we summarise the values and errors of the fit parameters in table \eqref{tbl:parameters}.
\begin{table}[htbp]
  \centering
  \begin{tabular}{|c||cc|}\hline \hline
           &       Estimate            &     Standard Error      \\ \hline \hline
   $A_1$   &  $3.61\times10^{-3}$      &   $9.80\times10^{-5}$         \\
   $A_2$   &  $4.50\times10^{-2}$      &   $6.73\times10^{-3}$         \\
   $A_3$   &  $3.42\times10^{-3}$      &   $5.69\times10^{-4}$         \\
   $A_4$   &  $1.11\times10^{-2}$      &   $2.26\times10^{-3}$         \\
   $B_1$   &  $5.37\times10^{-5}$      &   $3.55\times10^{-7}$         \\ 
   $B_2$   &  $9.11\times10^{-6}$      &   $4.80\times10^{-7}$         \\
   $B_3$   &  $1.20\times10^{-5}$      &   $4.87\times10^{-7}$         \\
   $B_4$   &  $8.71\times10^{-6}$      &   $4.95\times10^{-7}$         \\
   $C$     &  $-4.63\times10^{-1}$     &   $7.00\times10^{-2}$         \\
   $D$     &  $6.66\times10^{-1}$      &   $2.26\times10^{-2}$         \\   \hline
 \end{tabular}
 \caption{Fit parameters and errors respectively leading to the fitted curve in Fig.\eqref{fig:ExpReeFit}.}
 \label{tbl:parameters}
\end{table}
To perform the fit we used the NonlinearModelFit package of Mathematica software obtaining a $R^2$ value for the number of adjusted points equal to $R^2=0.9921$. A quite detailed discussion about the errors
appearing in the analysis of the experimental data is already presented in the second work of Ref.\cite{Matt} .

In order to compute the smeared experimental data we employ the same approach as the one of Ref.\cite{Matt}. We assume that the narrow resonances have a Breit-Wigner form
\be
\label{eq:BreitWigner}
R_{\mathrm{res}}(Q^2)=\frac{9}{\alpha^2}B_{\ell \ell}B_h\frac{M^2\Gamma^2}{(Q^2-M^2)^2+M^2\Gamma^2},
\ee
where $\alpha$ is the QED coupling, and $M,\, \Gamma,\, B_{\ell\ell},\, B_h$ are the mass, width, lepton branching ratio, and hadronic branching ratio respectively.
The narrow resonances that were excluded from the fit can now be represented by a delta function through the follow transformation
\be
\label{eq:deltaApprox}
\frac{1}{(Q^2-M^2)^2+M^2\Gamma^2} \approx \frac{\pi}{M\Gamma}\delta(Q^2-M^2).
\ee
For these we can write the Eq.\eqref{eq:BreitWigner} as 
\begin{equation}
\label{eq:BreitWignerdelta}
R_{\mathrm{res}}(Q^2) \approx \frac{9\pi M\Gamma}{\alpha^2}B_{\ell \ell}B_h \delta(Q^2-M^2).
\end{equation}

In order, to compare both the theoretical and experimental data we used the ``smearing'' procedure applied by Poggio, Quinn, and Weinberg (PQW) in Ref.\cite{pqw}. There they defined the ``smeared'' ratio by
\begin{equation}
 \label{eq:PQW}
 \overline{R}_{PQW}(Q^2;\Delta)=\frac{\Delta}{\pi}\int_{0}^{\infty}ds^{\prime}\frac{R_{e^+ e^-}(\sqrt{s^{\prime}})}{(s^{\prime}-Q^2)^2+\Delta^2}.
\end{equation}
The smeared ratio could be written as \cite{pqw}
\be
\label{eq:pqw}
2i \overline{R}_{PQW}(Q^2;\Delta)=\Pi(Q^2+i\Delta)-\Pi(Q^2-i\Delta).
\ee
where $\Pi(z)$ is the vacuum-polarization amplitude.
The best choice of $\Delta$ is the smallest value that will smooth out any rapid variations in either the experimental or the theoretical $R_{e^+ e^-}$.  It turns out that this depends upon the energy region one is interested in.
Around charm threshold a $\Delta = 3$ GeV$^2$ or more is necessary, while in the lowest-energy region a $\Delta$ as small as 1 GeV$^2$ can be used \cite{Matt}.
The idea now is to apply this smearing to both the theoretical and experimental $R_{e^+ e^-}$'s and then compare the results.

To determine the theoretical smeared ratio $R_{e^+ e^-}$ we can solve the Eq.\eqref{eq:PQW} by numerical integration. In order to deal with the different thresholds we performed the numerical integration considering several intervals, from $m_u$ to
$Q_{\mathrm{max}}$ and over the range 0 to $m_u$ we considered $R_{e^+ e^-}=0$. We assumed $Q_{\mathrm{max}}=6$ GeV.  From $Q_{\mathrm{max}}$ to $\infty$,  $R_{e^+ e^-}$ remained constant, and in this region we can make the integration of Eq.\eqref{eq:PQW}
analytically. The quite narrow resonances are given by Eq.\eqref{eq:BreitWignerdelta}, therefore the Eq.\eqref{eq:PQW} can be solved without trouble, and their contributions become
\be
\label{eq:SmearRres}
\overline{R}_{\mathrm{res}} \approx \frac{9B_{\ell\ell}B_h \Delta M\Gamma}{\alpha^2\big[(Q^2-M^2)^2+\Delta^2\big]}.
\ee
The other resonances ($\omega,\,\rho^{\prime},\,\psi(4040),\,\psi(4415)$)  were considered in the fit with a Breit-Wigner function.

\begin{figure}[htbp]
\setlength{\epsfxsize}{0.9\hsize} \centerline{\epsfbox{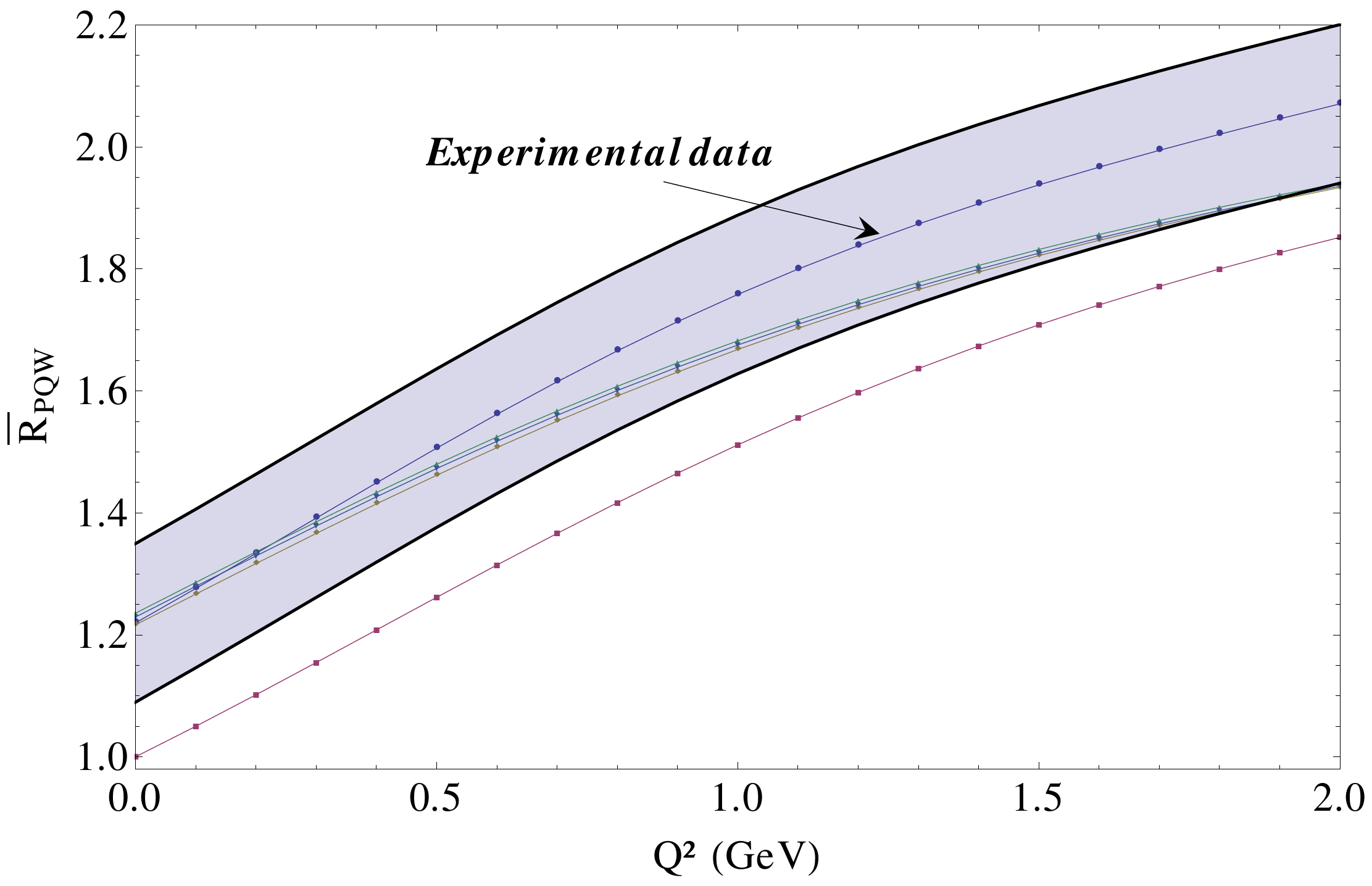}}
\caption[dummy0]{Smeared $R_{e^+ e^-}$ for $\Delta = 1.5$ GeV$^2$. The experimental result is shown in the middle of the shaded region. We have drawn this
shaded area assuming an error of $7\%$ above and below the smeared experimental data, what helped us to determine different
values of the ratio $m_g/\Lambda$ that are near this region. For example, the theoretical smeared $R_{e^+ e^-}$ for $m_g/\Lambda=0.7$ is shown by the magenta line, which is out
of the shaded area.}
  \label{fig:RpqwD15Ca}
\end{figure}

The fitted experimental data was  integrated numerically as shown in Eq.\eqref{eq:PQW}.
We computed the ``smeared'' quantity for four different values  $\Delta=$1 GeV$^2$, 1.5 GeV$^2$, 2 GeV$^2$, and 3 GeV$^2$. For simplicity we present in Fig.\eqref{fig:RpqwD15Ca} the result only for  $\Delta=1.5$ GeV$^2$.
The smeared theoretical $R_{e^+ e^-}$  was computed, threshold by threshold,
for values of $m_g/\Lambda$ between $0.7$ and $2.4$ and $\rho = 4$, and the values are also shown in Fig.\eqref{fig:RpqwD15Ca}.  Note that the QCD coupling constant Eq.\eqref{eq:alpha} depends on  $m_g/\Lambda$,
and consequently the ratio $R_{e^+ e^-}$ has the same dependence.
The shaded area in the figure was determined assuming $\pm 7 \, \%$ uncertainty 
around the experimental central value, obtained when we considered ad hoc variations of the many parameters in our fitting procedure and the possible normalisation errors pointed out in Ref.\cite{Matt}. 

In the Fig.\eqref{fig:RpqwD15Ca}  most of the theoretical smeared lines are within the shaded area except for the result obtained
with $m_g/\Lambda=0.7$ (magenta line with square data points), which is out of the shaded region.  This fact will happens for any 
$m_g/\Lambda$ value smaller than $0.7$, which lead to large values of the coupling constant where the expansion in $R_{e^+e^-}(s)$ does not make sense anymore.
This mean that there is not agreement between  data and theory for $m_g/\Lambda=0.7$ and lower values within the assumed uncertainty. 
The theoretically smeared $R_{e^+e^-}$  with $m_g/\Lambda$ between $0.8$ and $2.4$ are within the shaded region.  However,
larger $m_g/\Lambda$ values lead to theoretical $R_{e^+e^-}$ curves increasingly away from the experimental data, but with lines that barely can be distinguished from each other and we did not include these results in the figure.

We have applied the smearing method for the ratio $R_{e^+e^-}$ with $\rho=4$ and different values of $\Delta$ in order to compare experiment and theory. In principle, following Ref.\cite{Matt}, a smaller $\Delta$ value is better
to describe the low energy limit, and we shall limit ourselves to $R_{e^+e^-}$ up to $1.5$ GeV$^2$ as shown in Fig.\eqref{fig:RpqwD15Ca}. In order to compare the data we performed a $\chi^2$ test for our physical quantities. 
We choose a set of $\{\mathcal{O}_i^{exp}\}$ of measured observables and compute $\{\mathcal{O}_i^{th}(\bf{\theta})\}$, where $\bf{\theta}$ are parameters.  Then we minimise the $\chi^2$ function
$$\chi^2({\bf{\theta}})=\sum_i \frac{\big(\mathcal{O}_i^{exp}-\mathcal{O}_i^{th}({\bf{\theta}})\big)^2}{\big(\Delta \mathcal{O}_i^{exp}\big)^2},$$
where the observable of our interest is the ratio $R_{e^+e^-}$, and the parameter is the dynamic gluon mass $m_g/\Lambda$.  With 21 points 
(degrees of freedom) and a deviation of data about $10\%$ we develop a $\chi^2$ test for comparison of the smeared 
$R_{e^+e^-}$ at $\mathcal{O}(\alpha_s^3)$. 
In the Table \eqref{tbl:chi2} we set $\rho=4$ and write the $\chi^2$ values for several $m_g/\Lambda$. The values of $m_g/\Lambda$ are in the interval [0.8, 2.4], and according to Fig.\eqref{fig:RpqwD15Ca} they fall into the shaded region.

\begin{table}[htbp]
  \centering
  \begin{tabular}{|c||ccc|}\hline \hline
         $   $        &   $\Delta=1.5$    &    $\Delta=2$    &   $\Delta=3$    \\ \hline \hline
   $m_g/\Lambda = 0.8$   &    $1.37763$      &    $1.85683$     &    $2.70305$          \\
   $m_g/\Lambda = 0.9$   &    $0.837894$     &    $1.24134$     &    $2.10877$          \\
   $m_g/\Lambda = 1.0$   &    $0.712088$     &    $1.06814$     &    $1.92454$          \\
   $m_g/\Lambda = 1.2$   &    $0.657517$     &    $0.98275$     &    $1.82944$          \\
   $m_g/\Lambda = 1.4$   &    $0.666606$     &    $0.99339$     &    $1.84067$          \\ 
   $m_g/\Lambda = 1.6$   &    $0.698763$     &    $1.03745$     &    $1.88845$          \\
   $m_g/\Lambda = 1.8$   &    $0.744845$     &    $1.09813$     &    $1.95303$          \\
   $m_g/\Lambda = 2.0$   &    $0.801059$     &    $1.16891$     &    $2.02676$          \\
   $m_g/\Lambda = 2.2$   &    $0.865123$     &    $1.24620$     &    $2.10588$          \\
   $m_g/\Lambda = 2.4$   &    $0.935314$     &    $1.32809$     &    $2.18823$          \\   \hline
 \end{tabular}
 \caption{$\chi^2$-values for $R_{e^+e^-}$ with $\rho=4$ and different $\Delta$ values (in units of GeV$^2$).}
 \label{tbl:chi2}
\end{table}

\begin{figure}[htbp]
\setlength{\epsfxsize}{1.0\hsize} \centerline{\epsfbox{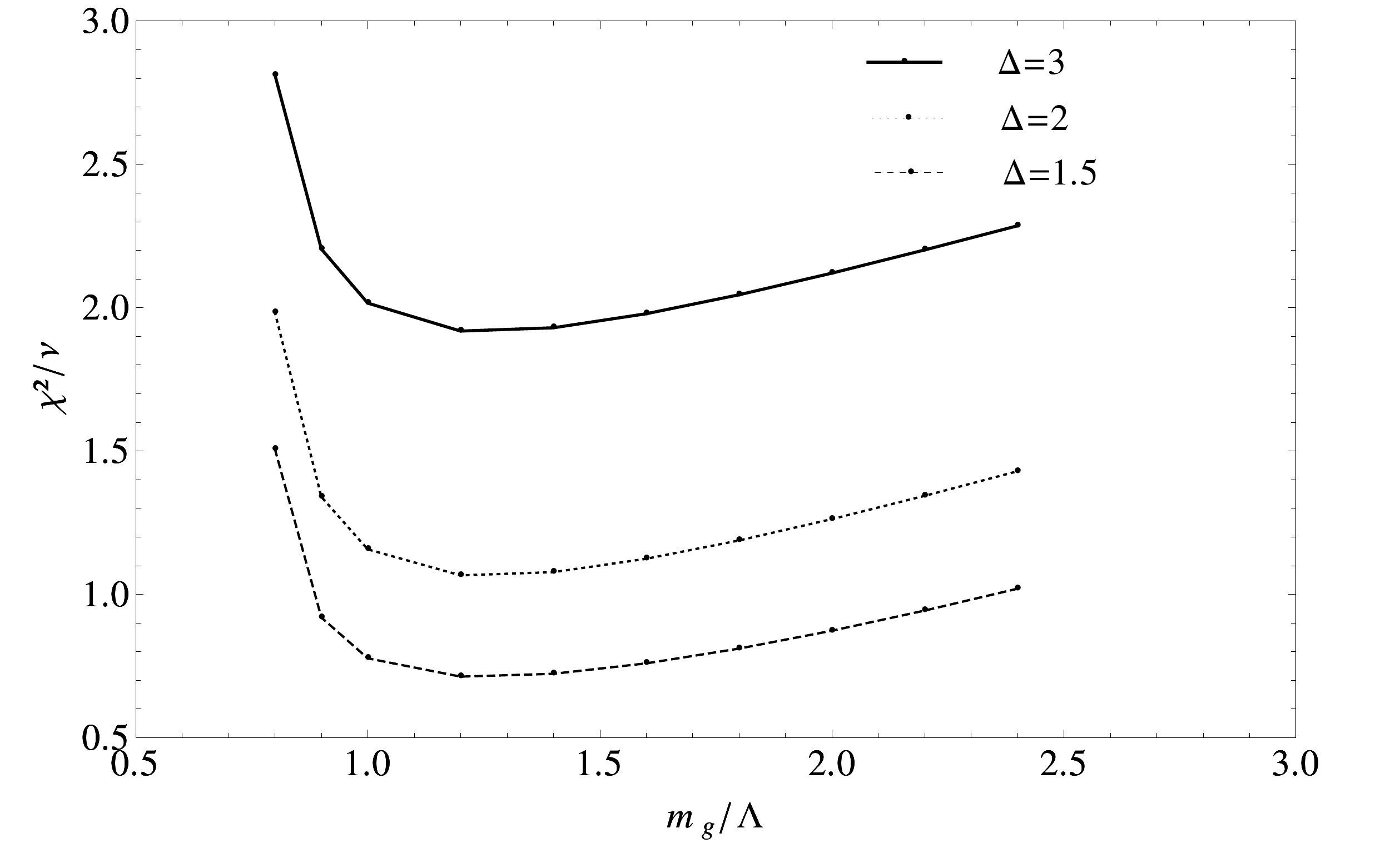}}
\caption[dummy0]{$\chi^2$ for the smeared quantity $\overline{R}_{e^+e^-}(Q^2,\Delta)$ with $\rho=4$ and different $\Delta$ values.}
  \label{fig:Chi2rho4}
\end{figure}

One interesting point in the table \eqref{tbl:chi2} is that for different $m_g/\Lambda$ values and $\rho=4$ all minimum values of $\chi^2$  occur for $m_g/\Lambda=1.2$.
This behaviour can be better seen in Fig.\eqref{fig:Chi2rho4}, where $\nu$ is the number of degrees of freedom.
This means that { \textit{there is one value of $m_g/\Lambda$ that provides the best match with the experiment}}, and points out
for one specific infrared value of the QCD coupling constant. It is important to remember that in the SDE equations $m_g$ is one input parameter and lattice data has not enough precision to pinpoint this mass scale,
revealing the interest on phenomenological determinations of this quantity.

\begin{figure}[ht!]
\setlength{\epsfxsize}{1.0\hsize} \centerline{\epsfbox{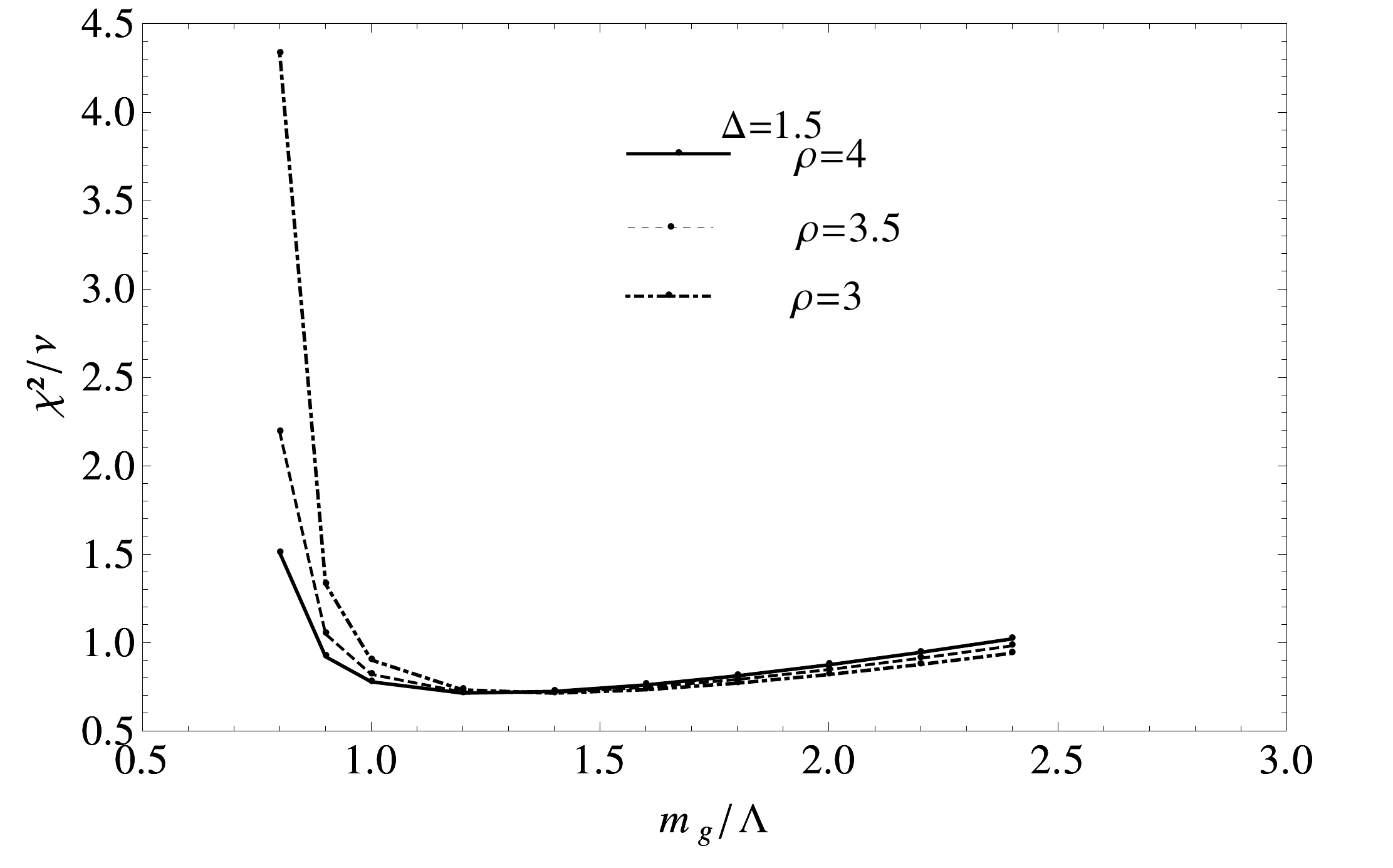}}
\caption[dummy0]{$\chi^2$ for the smeared quantity $\overline{R}_{e^+e^-}(Q^2,\Delta)$ with $\Delta = 1.5$ GeV$^2$ and different  
$\rho$ values.}
  \label{fig:Chi2D15}
\end{figure}

\begin{figure}[ht!]
\setlength{\epsfxsize}{1.0\hsize} \centerline{\epsfbox{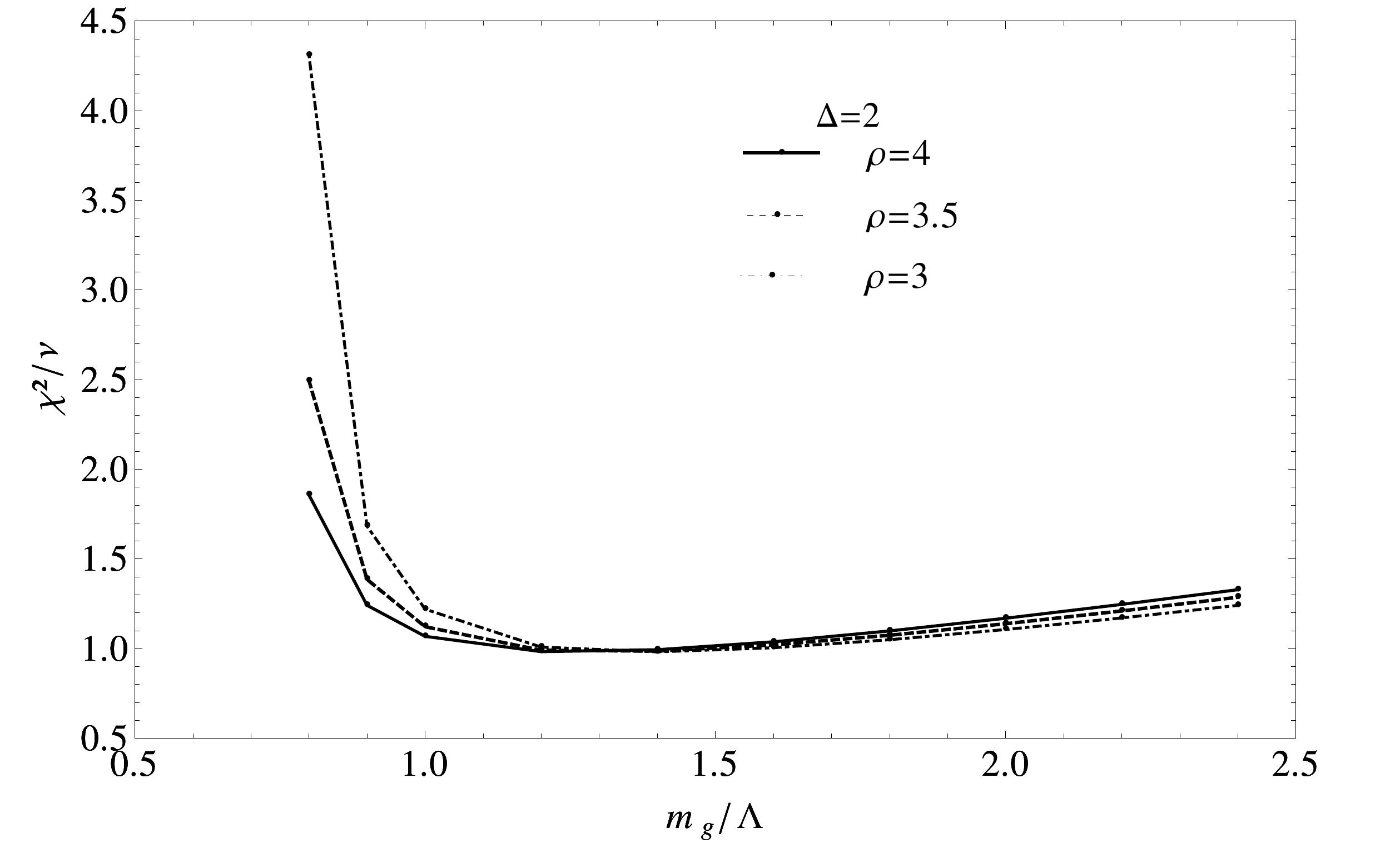}}
\caption[dummy0]{$\chi^2$ for the smeared quantity $\overline{R}_{e^+e^-}(Q^2,\Delta)$ with $\Delta = 2$ GeV$^2$ and different  
$\rho$ values.}
  \label{fig:Chi2D2}
\end{figure}

As we discussed after Eq.(\ref{eq:2}) for each value of $m_g/\Lambda$ we have a slightly different SDE solution, whose differences
can be parameterised in terms of different $\rho$ values. Therefore we have changed the $\Delta$ values, the $\rho$ values and
studied the $\chi^2$ distribution. With $\rho=4$ and $\Delta=1.5,\,2,\,3$ GeV$^2$ we have found that the minimum is located at
 $m_g/\Lambda=1.2$. However, with $\Delta=1.5$ GeV$^2$, $\Delta=2$ GeV$^2$ and taking $\rho=3,3.5,4$, we can see in Fig.\eqref{fig:Chi2D15} and Fig.\eqref{fig:Chi2D2} that the minimum points are now located between $m_g/\Lambda=1.2$ and $m_g/\Lambda=1.4$. Note that the minimum 
for $\Delta=(1.5,\, 2$) GeV$^2$ and $\rho=3,\,3.5$ are approximately at $m_g/\Lambda=1.4$. Unless some SDE solution deviates
grossly from the results that are presently found in the literature, $m_g/\Lambda$ values between $1.2$ and $1.4$ and $\rho$ values
between $3$ and $4$ lead to an infrared value of the coupling constant, according to Eq.(\ref{eq:alpha}) with two quark flavors, approximately of $\mathcal{O} ( 0.7)$. The minimum $\chi^2$ seems to be dependent on the
product $\rho m_g/\Lambda$.

\section{Global duality and the infrared QCD coupling}

It is usually assumed that there is a frontier between perturbative and non-perturbative QCD at one specific mass scale, and there are proposals about how this scale, that separates long and short distances, can be determined.
One of these proposals is established by the concept of global
duality and is explained in Refs.\cite{dual1,dual2}.  The determination of this scale ($s_0$) consist in doing a matching between perturbative QCD (pQCD) and experiment by the
expression
\be
\int_{0}^{s_0} dt \frac{1}{\pi} Im \, \Pi (t)_{exp}=\int_{0}^{s_0} dt \frac{1}{\pi} Im \, \Pi (t)_{pQCD}.
\label{la1}
\ee
The $s_0$ scale was determined to be of $s_0\approx 1.5$ GeV$^2$  \cite{dual2,milton}.

Eq.(\ref{la1}) also imply 
\begin{equation}
\int_{0}^{s_0} ds \, \overline{R}_{exp}=\int_{0}^{s_0} ds \, \overline{R}_{theor},
\label{eq:duality}
\end{equation}
and we can use our previous result to determine  $s_0$. It is clear that we shall have a different $s_0$ value for the matching prescribed by Eq.\eqref{la1} if we use
the effective charge that we discussed before. This happens because in the case of perturbative QCD the right-hand side of Eq.(\ref{la1}) starts growing fast as we
approach the Landau pole of the perturbative coupling and the solution appears at a large mass scale. On the other hand assuming global duality for the
IR finite effective QCD charge we do not expect a fast rise of the coupling at low momenta and consequently a slow increase of the right-hand side of Eq.(\ref{eq:duality}) as we go to small momenta.

To determine the scale  $s_0$ we can define the quantity:
\be
\epsilon(Q^2,s_0)=\int_{0}^{s_0} ds \, (\overline{R}_{exp}-\overline{R}_{theor}).
\label{eq:epsdual}
\ee
and look for zeros of this equation. In
the above equation we have a $s_0$ and $m_g/\Lambda$ dependence besides the one on $\Delta$ and $\rho$. 
For simplicity we can solve Eq.(\ref{eq:epsdual}) in the limit $Q^2\rightarrow 0$, for specific values of the smearing factor $\Delta$, $\rho$ and $m_g/\Lambda$  and look for a 
zero of Eq.(\ref{eq:epsdual}).

\begin{figure}[htbp]
\setlength{\epsfxsize}{1.0\hsize} \centerline{\epsfbox{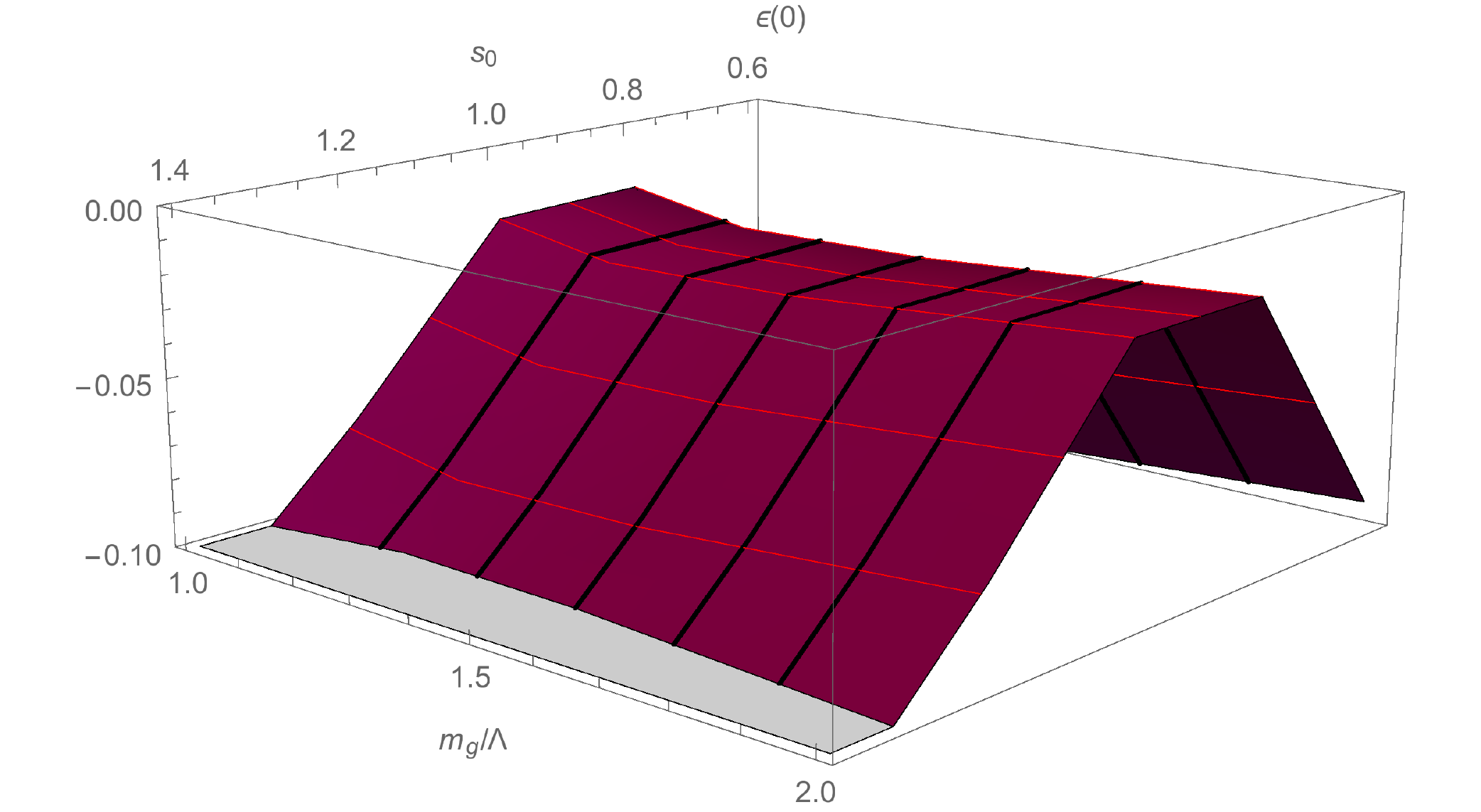}}
\caption[dummy0]{$\epsilon(Q^2 \rightarrow 0)$ surface (Eq.\eqref{eq:epsdual}) calculated up to $\mathcal{O}(\alpha_s^3)$ with $\Delta=1.5$ and $\rho =4$.}
  \label{fig:DualityD15a}
\end{figure}

In the case of $\Delta = 1.5$ GeV$^2$, $\rho =4$ and $m_g/\Lambda = 1.2$ we show in Table \eqref{tbl:dual} that $\epsilon (Q^2 \rightarrow 0) \approx 0$ when $s_0 = 0.87$ GeV$^2$.
This, as discussed previously, is the result that we were expecting when we change the perturbative coupling by the effective charge of Eq.(\ref{eq:alpha}) , showing that the so called
frontier between the perturbative and non-perturbative physics occurs at one smaller mass scale than the one determined with perturbative QCD. We also show in Fig.\eqref{fig:DualityD15a}
a plot of  $\epsilon (Q^2 \rightarrow 0)$ as a function of $s_0$ and $m_g/\Lambda $ in the case of $\Delta =  1.5$ GeV$^2$ and $\rho =4$, where we can show that the smallest $\epsilon (0)$ values
are obtained at one $s_0$ smaller than $1.5$ GeV$^2$.  Actually the effective coupling that we are using has non-perturbative
information, therefore we should not be allowed to say that $s_0$ is the mass scale indicating the transition between short and large distances, but just say that if we improve the
QCD calculations with such effective charge we probably could perform ``improved perturbative" calculations a little bit deeper into the IR region.

\begin{table}[htbp]
  \centering
  \begin{tabular}{|c|c|}\hline
    $s_0$   &   $\epsilon(0)$      \\ \hline \hline
   $ 0.6$   &    $-0.0987$              \\
   $ 0.8$   &    $-0.0276$              \\
   $ 0.87$  &    $-0.0060$               \\
   $ 1.0$   &    $-0.0291$               \\
   $ 1.2$   &    $-0.0812$               \\
   $ 1.4$   &    $-0.1264$              \\
   $ 1.6$   &    $-0.1545$              \\  \hline
 \end{tabular}
 \caption{$\epsilon(0)$ with $m_g/\Lambda = 1.2$, $\rho =4$ and different $s_0$ values.}
 \label{tbl:dual}
\end{table}

\section{Conclusions}

We have studied the electron-positron annihilation process into hadrons $R_{e^+e^-}$ up to $\mathcal{O}(\alpha_{s}^{3})$, where the QCD coupling constant was described by an effective charge obtained in solutions of the Schwinger-Dyson
equations in the pinch technique scheme. This effective charge is frozen in the infrared region and the frozen value is related to a dynamically generated gluon mass. The existence of such dynamical gluon mass, or an effective mass
scale for the gluon propagator, has not only been observed in SDE solutions but has been confirmed by lattice simulations, and in this case we can prove that the coupling constant freezes in the IR region \cite{us1}. The
main purpose of the work was the determination of the infrared value of the dynamical gluon mass, related to the IR value of the effective charge, since this value enters as an input into the numerical solutions of the SDE, and lattice data including dynamical
quarks can only obtain rough approximations to this quantity.

To compare the $R_{e^+e^-}$ experimental data to the theoretical calculation we adopted the smearing method suggest by Poggio, Quinn and Weinberg.
 In order to find the best fit between experimental data and theory, we performed a $\chi^2$ study of the resulting curves,
that, within the uncertainties of the approach, leads to a minimum value when $m_g/\Lambda_{QCD}$ is in the range $1.2 \, - \, 1.4$. These values are in agreement with other phenomenological determinations of this ratio and imply
an infrared effective charge $\alpha_s(0) \approx 0.7$, what is also in agreement with the result obtained by Mattingly and Stevenson when analysing $R_{e^+e^-}$  in a different scheme \cite{Matt}.

The $\chi^2$ study indicates that the result is stable and independent of the smearing parameter $\Delta$. All uncertainties about the full procedure have been discussed in Ref.\cite{Matt}, and the main novelty is the use
of the effective charge related to the dynamical gluon mass scale.  Although there are many evidences for an IR finite gluon propagator and coupling constant, there is not an unique definition of the  non-perturbative
QCD charge. However, the effective coupling that we have considered here can map any possible behaviour of the QCD charge, i.e. is compatible with the UV behaviour predicted by perturbative QCD and freezes in
the IR region as should be expected when the theory develops a dynamically generated mass. 

Taking advantage of our $R_{e^+e^-}$ calculation we have discussed what happens with the scale determined by the concept of global duality \cite{dual1,dual2} when the perturbative coupling is exchanged by the effective charge discussed up to now.
We verify that $s_0$, the mass scale indicating the transition between short and large distances, is changed to a smaller value. However, as should be understood from our procedure, we are assuming an ``improved" coupling,
and this fact has clear phenomenological consequences. One is that we could perform ``improved perturbative" calculations possibly going deeper into the IR region, although the main point is that it is necessary to build a bridge between
perturbative QCD and the SDE results, which not only show a cure for the QCD IR divergences, but seems to provide a soft transition between the perturbative and non-perturbative QCD regimes.

Although the idea of dynamical gluon mass generation was put forward a long time ago, leading to infrared
finite Green's functions \cite{cornwall}, due to the involved field theoretical aspects of the combination of the pinch technique with the
background field method necessary to obtain gauge invariant and renormalization group independent quantities, only after
the appearance of many lattice QCD simulations this subject had a revival, in such a way that very recent theoretical papers
are revisiting the determination of a finite gluon propagator and coupling constant in three \cite{ne54} and four dimensional
QCD \cite{ne41}, trying to spread the ideas of this mechanism. On the other hand this subject also raised attempts to describe these infrared finite quantities in terms of an
effective Lagrangian valid for infrared QCD (see Refs.\cite{ne55,ne56,ne57} and references therein). These attempts, although introducing a
hard gluon mass and leading to a ``soft" BRST symmetry breaking, seem to be compatible (at the ``perturbative" level) with many lattice data about the freezing
of the infrared QCD quantities. Therefore, including the discussion presented in this work, there are several new \cite{ne58,ne59,ne60,ne61} 
and old \cite{ne62} tests for the phenomenological consequences of an infrared
finite QCD coupling and gluon propagator, and, as the time goes on, all these attempts may provide a new consistent view of infrared QCD. It is also interesting to note that in a recent work \cite{ne63} it was determined the transition scale between perturbative and non-perturbative QCD using Hadron-Parton duality, which, when translated to the ${\overline{MS}}$ scheme, would give a value close to the  coupling determined by us. This can only indicate that the different approaches lead to a result that is quite stable.

\vspace{0.5cm}
\section*{Acknowledgments}
\vspace{-0.5cm}
We have befited from discussions with D. A. Fagundes. This research was partially supported
by the Conselho Nac. de Desenv. Cient\'{\i}fico e Tecnol\'ogico
(CNPq), by the grants 2013/22079-8 and 2013/24065-4 of Funda\c c\~ao de Amparo \`a Pesquisa do
Estado de S\~ao Paulo (FA\-PES\-P) and by Coordena\c c\~ao de Aper\-fei\-\c coa\-mento
de Pessoal de N\'{\i}vel Superior (CAPES).


 \end{document}